\documentclass[conference]{IEEEtran}
\IEEEoverridecommandlockouts
\usepackage{cite}
\usepackage{amsmath,amssymb,amsfonts}
\usepackage{algorithmic}
\usepackage{graphicx}
\usepackage{textcomp}
\usepackage{xcolor}
\def\BibTeX{{\rm B\kern-.05em{\sc i\kern-.025em b}\kern-.08em
    T\kern-.1667em\lower.7ex\hbox{E}\kern-.125emX}}
    
\usepackage{glossaries}
\usepackage{multicol}
\usepackage{adjustbox}
\usepackage{multirow}
\usepackage{siunitx}
\usepackage{bm}
\usepackage{booktabs}
\usepackage{xspace}

\usepackage{microtype}  

\usepackage{cite}
\usepackage{amsmath,amssymb,amsfonts}
\usepackage{algorithmic}
\usepackage{graphicx}
\usepackage{textcomp}
\usepackage{glossaries}
\usepackage{xcolor}
\usepackage{graphicx}
\usepackage{psfrag}
\usepackage{siunitx}
\usepackage{bm,bbm}
\usepackage{mathtools}
\usepackage{booktabs}
\usepackage[capitalise]{cleveref}
\usepackage{url}
\usepackage{fontawesome}

\usepackage{amssymb}
\usepackage{pifont}

\usepackage{booktabs}
\usepackage{mathtools}
\usepackage{xcolor}
\usepackage{tikz,pgfplots}
\usepackage{tikzscale}
\usepgfplotslibrary{groupplots}
\usetikzlibrary{arrows, arrows.meta, decorations.markings}

\usepackage{tikz}
\usepackage{standalone}
\usepackage{pgfplots}

\usetikzlibrary{positioning,chains,calc,shapes.geometric, shadows, shapes.misc, fit, arrows, shapes.callouts}
\usepgfplotslibrary{groupplots}
\usetikzlibrary{backgrounds}

\definecolor{yellow}{RGB}{255, 198, 0}
\definecolor{orange}{RGB}{255, 130, 0}
\definecolor{blue}{RGB}{0, 32, 91}
\definecolor{red}{RGB}{198, 53, 39}
\definecolor{magenta}{RGB}{138, 27, 97}
\definecolor{lightblue}{RGB}{0, 159, 223}
\definecolor{green}{RGB}{0, 155,119}
\definecolor{lightgreen}{RGB}{132, 189,0}
\definecolor{greenyellow}{RGB}{208, 223,0}

\newacronym{DNN}{DNN}{deep neural network}
\newacronym{NN}{NN}{neural network}
\newacronym{CDR}{CDR}{coherent-to-diffuse power ratio}
\newacronym{DoA}{DOA}{direction of arrival}
\newacronym[longplural={time differences of arrival}]{TDoA}{TDOA}{time difference of arrival}
\newacronym{MSE}{MSE}{mean-squared error}
\newacronym{MSD}{MSD}{mean-squared deviation}
\newacronym{MLP}{MLP}{multilayer perceptron}
\newacronym{CRNN}{CRNN}{convolutional recurrent neural network}
\newacronym[plural=GPs,firstplural=Gaussian processes (GPs)]{GP}{GP}{Gaussian process}
\newacronym{GRU}{GRU}{gated recurrent unit}
\newacronym{RIR}{RIR}{room impulse response}
\newacronym{AWGN}{AWGN}{additive white Gaussian noise}
\newacronym{SNR}{SNR}{signal-to-noise ratio}
\newacronym{STFT}{STFT}{short-time Fourier transform}
\newacronym[plural=PSDs,firstplural=power spectral densities~(PSDs)]{PSD}{PSD}{power spectral density}
\newacronym{AE}{AE}{absolute error}
\newacronym{MAE}{MAE}{mean-absolute error}
\newacronym{PE}{PE}{position error}
\newacronym{MPE}{MPE}{mean position error}
\newacronym{WASN}{WASN}{wireless acoustic sensor network}
\newacronym{OoR}{OoR}{out-of-range}
\newacronym{ASR}{ASR}{automatic speech recognition}
\newacronym{TDNN}{TDNN}{time delay neural network}
\newacronym{CNN}{CNN}{convolutional neural network}
\newacronym{WLS}{WLS}{weighted least squares}
\newacronym{LS}{LS}{least squares}
\newacronym{RANSAC}{RANSAC}{random sample consensus}
\newacronym{MVDR}{MVDR}{minimum variance distortionless response}

\newacronym{GARDE}{GARDE}{\textbf{G}eometry c\textbf{A}libration f\textbf{R}om \textbf{D}istance \textbf{E}stimates}
\newacronym{MDS}{MDS}{Multi Dimensional Scaling}

\newacronym{CWLS}{CWLS}{constrained weighted least squares}
\newacronym{CRLB}{CRLB}{Cramer-Rao lower bound}
\newacronym{RMSE}{RMSE}{root mean square error}

\newacronym{CDF}{CDF}{cumulative distribution function}
\newacronym{CDF2}{CDF}{Cumulative distribution function}

\newacronym{ABEL}{ABEL}{averaged burst error length}
\newacronym{ACD}{ACD}{average coherence drift}
\newacronym{ADC}{ADC}{analog-digital converter}
\newacronym{APT}{APT}{averaged processing time}
\newacronym{ASN}{ASN}{acoustic sensor network}
\newacronym{ASNs}{ASNs}{acoustic sensor networks}
\newacronym{ASRC}{ASRC}{arbitrary sampling rate conversion}
\newacronym{ATD}{ATD}{time drift}
\newacronym{ATS}{ATS}{accumulating time shift}
\newacronym{BI}{BI}{band-limited interpolation}
\newacronym{BSS}{BSS}{blind source separation}
\newacronym{CCF}{CCF}{cross-correlation function}
\newacronym{CCF-2}{CCF-2}{secondary \gls{CCF}}
\newacronym{CD}{CD}{coherence drift}
\newacronym{CFM}{CFM}{coherence function maximization}
\newacronym{CM}{CM}{correlation maximization}
\newacronym{CSD}{CSD}{cross-spectral density}
\newacronym{CSD-2}{CSD-2}{secondary \gls{CSD}}
\newacronym{CTC}{CTC}{continuous-time conversion}
\newacronym{DCML}{DCML}{\gls{ML} method for dynamic conditions}
\newacronym{DB}{DB}{Data base}
\newacronym{DD}{DD}{digital-to-digital}
\newacronym{DDC}{DDC}{digital-to-digital converter}
\newacronym{DSP}{DSP}{digital signal processing}
\newacronym{DTC}{DTC}{discrete-time conversion}
\newacronym{DXCP}{DXCP}{double-cross-correlation processor}
\newacronym{DXCPPhaT}{DXCP-PhaT}{\gls{DXCP} with phase transform}
\newacronym{ECS}{ECS}{energy correlation score}
\newacronym{FCF}{FCF}{filter correlation function}
\newacronym{FF}{FF}{free-field}
\newacronym{FFT}{FFT}{fast Fourier transform}
\newacronym{FFT-DXCP}{FFT-DXCP}{\gls{FFT} domain \gls{DXCP}}
\newacronym{FO}{FO}{frame-oriented}
\newacronym{GCC}{GCC}{generalized cross-correlation}
\newacronym{GCPSD}{GCPSD}{generalized cross power spectral density}
\newacronym{GCCF}{GCCF}{generalized \gls{CCF}}
\newacronym{GCCPhaT}{GCC-PhaT}{generalized cross-correlation with phase transform}
\newacronym{GCSD}{GCSD}{generalized cross-spectral density}
\newacronym{GE}{GE}{Gilbert-Elliott}
\newacronym{IBI}{IBI}{iterative \gls{BI}}
\newacronym{ICA}{ICA}{independent component analysis}
\newacronym{IML}{IML}{iterative \gls{ML}}
\newacronym{IR}{IR}{impulse response}
\newacronym{IIR}{IIR}{infinite impulse response}
\newacronym{IFFT}{IFFT}{inverse fast Fourier transform}
\newacronym{ISTFT}{ISTFT}{inverse \gls{STFT}}
\newacronym{LCD}{LCD}{least-squares coherence drift}
\newacronym{LPD}{LPD}{linear-phase drift}
\newacronym{LTI}{LTI}{linear time-invariant}
\newacronym{LTV}{LTV}{linear time-variant}
\newacronym{LTVIR}{LTV-IR}{linear time-variant impulse response}
\newacronym{ML}{ML}{maximum likelihood}
\newacronym{MS}{MS}{multi-stage}
\newacronym{MSC}{MSC}{magnitude-squared coherence}
\newacronym{OML}{OML}{optimized \gls{ML}}
\newacronym{OR}{OR}{outlier removal}
\newacronym{PHAT}{PHAT}{phase transform}
\newacronym{PL}{PL}{packet loss}
\newacronym{PolyFar}{PolyFar}{polyphase-Farrow}
\newacronym{ppb}{ppb}{part per billion}
\newacronym{ppm}{ppm}{parts \ per \ million}

\newacronym{RBI}{RBI}{recursive band-limited interpolation}
\newacronym{RF}{RF}{realtime factor}
\newacronym{RTP}{RTP}{Real-time Transport Protocol}
\newacronym{RTCP}{RTCP}{Real-time Transport Control Protocol}

\newacronym{SCOT}{SCOT}{smoothed coherence transform}
\newacronym{SINR}{SINR}{signal-to-interpolation-noise ratio}
\newacronym{SO}{SO}{sample-oriented}
\newacronym{SPIB}{SPIB}{signal processing information base}
\newacronym{SPL}{SPL}{signal packet loss}
\newacronym{SRC}{SRC}{sampling rate conversion}
\newacronym{SRO}{SRO}{sampling rate offset}
\newacronym[longplural={spatial covariance matrices}]{SCM}{SCM}{spatial covariance matrix}
\newacronym{STO}{STO}{sampling time offset}
\newacronym{TCP}{TCP}{Transmission Control Protocol}
\newacronym{TD}{TD}{time domain}
\newacronym{TDDXCP}{TD-DXCP}{time domain \gls{DXCP}}
\newacronym{TDD}{TDD}{time-delay difference}
\newacronym{UDP}{UDP}{User Datagram Protocol}
\newacronym{VAD}{VAD}{voice activity detection}
\newacronym{SAD}{SAD}{sound activity detection}
\newacronym{WACD}{WACD}{weighted average coherence drift}
\newacronym{DWACD}{DWACD}{dynamic weighted average coherence drift}
\newacronym{WG}{WG}{weighting}
\newacronym{WLCD}{WLCD}{weighted \gls{LCD}}
\newacronym{WASNs}{WASNs}{wireless acoustic sensor networks}
\newacronym{WCP}{WCP}{wideband correlation processor}
\newacronym{XC}{XC}{cross-correlation}
\newacronym{SVD}{SVD}{singular value decomposition}

\newacronym{TXCO}{TXCO}{temperature compensated crystal oscillator}

\newacronym{OU}{OU}{Ornstein-Uhlenbeck}

\newacronym[firstplural=time differences of flight (TDOFs)]{TDOF}{TDOF}{time difference of flight}

\newacronym{cACGMM}{cACGMM}{complex Angular Central Gaussian Mixture Model}
\newacronym{EM}{EM}{Expectation Maximization}
\newacronym{EVD}{EVD}{eigenvalue decomposition}

\newacronym{cpWER}{cpWER}{concatenated minimum-permutation word error rate}
\newacronym{WER}{WER}{word error rate}

\newacronym{SDR}{SDR}{signal-to-distortion ratio}

\newacronym{SRP}{SRP}{steered response power}
\newacronym{SRPPhaT}{SRP-PhaT}{steered-response power with phase transform}

\newacronym{GSS}{GSS}{guided source separation}
\newacronym{WPE}{WPE}{weighted prediction error}
\newacronym{EEND}{EEND}{End-to-End Neural Diarization}

\begin{document}
\setlength{\textfloatsep}{4pt}
\setlength{\intextsep}{4pt}
\setlength{\intextsep}{2pt}
\setlength{\abovedisplayskip}{2pt}
\setlength{\abovedisplayshortskip}{2pt}
\setlength{\belowdisplayskip}{2pt}
\setlength{\belowdisplayshortskip}{2pt}
\clubpenalty = 10000
\widowpenalty = 10000
\displaywidowpenalty = 10000

\title{Spatial Diarization for Meeting Transcription \\ with Ad-Hoc Acoustic Sensor Networks}

\author{\IEEEauthorblockN{Tobias Gburrek, Joerg Schmalenstroeer  and Reinhold Haeb-Umbach}
\IEEEauthorblockA{\textit{Department of Communications Engineering} \\
\textit{Paderborn University, Germany}\\
\{gburrek, schmalen, haeb\}@nt.uni-paderborn.de}
}

\maketitle

\begin{abstract}
    We propose a diarization system, that estimates ``who spoke when'' based on spatial information, to be used as a front-end of a meeting transcription system running on the signals gathered from an \gls{ASN}.
    Although the spatial distribution of the microphones is advantageous, exploiting the spatial diversity for diarization and signal enhancement is challenging, because the microphones' positions are typically unknown, and the recorded signals are initially unsynchronized in general.
    Here, we approach these issues by first blindly synchronizing the signals and then estimating \glspl{TDoA}. 
    The \gls{TDoA} information is exploited to  estimate the speakers' activity, even in the presence of multiple speakers being simultaneously active. 
    This speaker activity information serves as a guide for a spatial mixture model, on which basis the individual speaker's signals are extracted via beamforming. 
    Finally, the extracted signals are forwarded to a speech recognizer.
    Additionally, a novel initialization scheme for spatial mixture models based on the \gls{TDoA} estimates is proposed.
    Experiments conducted on real recordings from the LibriWASN data set have shown that our proposed system is advantageous compared to a system using a spatial mixture model, which does not make use of external diarization information.
\end{abstract}


\section{Introduction}
\label{sec:intro}
    When transcribing a meeting, often not only the information  of what has been said is of interest but also the information ``who spoke when'', i.e., diarization information.
    Additionally, diarization information can also be helpful for speech enhancement, e.g., using the \gls{GSS}~\cite{PB2018CHiME5} framework.
    However, gathering diarization information is a challenging task due to the highly dynamic nature of spontaneous conversations with alternating silence and speech regions, as well as overlapping speech from multiple speakers. 
    
    In particular, the segments with overlapping speech are challenging for diarization. 
    For example, the performance of methods, that rely on spectro-temporal information, often tends to degrade with an increasing amount of overlapping speech.
    This especially holds for early diarization systems \cite{second_DIHARD}. 
    Although nowadays diarization systems, like TS-VAD~\cite{medennikov20_interspeech}, are able to cope much better with overlap, their performance is often still negatively affected by overlap~\cite{he21c_interspeech}.
    
    In a typical meeting scenario with multiple speakers sitting around a table at spatially well separated, (quasi-)fixed positions the information ``when and at which position'' a speaker is active also reveals the diarization information. 
    In such a scenario spatial information can be a promising alternative to cope with speech overlap. 
    Typically, \gls{DoA} information, which is gathered using a compact microphone array, is employed as source of spatial information~\cite{araki08, ishiguro12, fakhry16, zheng22, taherian2023multichannel}.
    
    Discriminating between two speakers based on \gls{DoA} information might be challenging, if the distance between the speakers and the microphone array is large and the speakers sit close to each other. 
    The spatial diversity of an \gls{ASN} comes in handy in such situations by offering \gls{TDoA} information, which allows for a better distinction between those speakers.
    However, \glspl{ASN} are typically formed ad-hoc, e.g., by smartphones.
    Hence, the microphone positions are generally unknown and the recorded signals are typically asynchronous, which makes it difficult to infer the speakers' position from the \gls{TDoA} estimates. 
    In~\cite{Gburrek22b} we approached these issues by using geometry calibration~\cite{GeoJournal} and a complex synchronization method~\cite{Gburrek22}, which maintains the information about the microphones’ and speakers’ positions, as preprocessing steps before diarization. 
    
    Here, a much simpler approach to synchronization is employed, which however distorts the information about the microphones’ and speakers’ positions by constant \gls{TDoA} offsets.
    Although, these distortions make it difficult to map the \glspl{TDoA} to the coordinates of the speakers' positions anymore, the \glspl{TDoA} still uniquely represent the speakers' positions.
    Hence, we propose to derive diarization information by clustering estimates stemming from a  multi-speaker \gls{TDoA} estimator, which delivers estimates at frame rate.
   
    The resulting diarization information is used as a guide for a spatial mixture model in the \gls{GSS} framework, to force the posterior probability to be zero when a speaker is inactive. 
    In experiments on the LibriWASN~\cite{schmalen23} data set we show that the guided spatial mixture model is able to outperform a blind spatial mixture model, which does not employ external diarization information.
    Additionally, a time-frequency bin wise initialization scheme for a spatial mixture-model based on the \gls{TDoA} estimates is proposed to speed up the convergence.

    In the following we describe the considered meeting scenario in~\cref{sec:scenario} and give an overview of the  meeting transcription pipeline in~\cref{sec:system}.
    Afterwards, the proposed \gls{TDoA}-based diarization system is introduced in~\cref{sec:diarization}, followed by a description, how the diarization information and the \gls{TDoA} estimates can be employed to support source extraction, in~\cref{sec:extraction}. 
    Experimental results are reported in~\cref{sec:exp}.
    Finally, conclusions are drawn in~\cref{sec:conclusions}. 

\section{Scenario Description}
\label{sec:scenario}
In the following a meeting-like conversation of $I$ speakers is considered, which should be transcribed.
It is assumed that the speakers sit at spatially well separated, fixed but unknown positions around a table. 
During the conversation, there are periods in time without speech activity, periods in time with a single speaker being active and a significant amount of periods in time with two speakers being active at the same time.
On the table, $M \geq 4$ microphones, forming an ad-hoc \gls{ASN}, are distributed, which are used to record the meeting.
The microphones are located at fixed but unknown positions.


Since the devices in an ad-hoc \gls{ASN} are generally independent of each other, the microphone signals are sampled with slightly different sampling frequencies even though the devices have the same nominal sampling rate. 
This introduces a \gls{SRO} between the microphone signals. 
Furthermore, the devices usually start their recordings at different points in time, which causes a \gls{STO} between the microphone signals.

\section{Meeting Transcription System}
\label{sec:system}
\begin{figure}[b]
    \centering
    \hspace{-1cm}
    \begin{adjustbox}{width=.75\columnwidth,center}
	    \tikzset{%
  block/.style    = {draw, thick, rectangle, minimum height = 2.7em, minimum width = 10.em, fill=white, align=center, rounded corners=0.1cm},
  block2/.style    = {draw, thick, rectangle, minimum height = 2.7em, minimum width = 13.50em, fill=white, align=center, rounded corners=0.1cm},
  block1/.style    = {draw, thick, rectangle, minimum height = 2.7em, minimum width = 8.25em, align=center, rounded corners=0.1cm},
  sum/.style      = {draw, circle, node distance = 2cm}, 
  cross/.style={path picture={\draw[black](path picture bounding box.south east) -- (path picture bounding box.north west)
		 (path picture bounding box.south west) -- (path picture bounding box.north east);}},
	   zigzag/.style = {
	   	to path={ -- ($(\tikztostart)!.55!-9:(\tikztotarget)$) --
	   		($(\tikztostart)!.45!+9:(\tikztotarget)$) -- (\tikztotarget)
	   		\tikztonodes},sharp corners}
               }
\tikzset{fit margins/.style={/tikz/afit/.cd,#1,
    /tikz/.cd,
    inner xsep=\pgfkeysvalueof{/tikz/afit/left}+\pgfkeysvalueof{/tikz/afit/right},
    inner ysep=\pgfkeysvalueof{/tikz/afit/top}+\pgfkeysvalueof{/tikz/afit/bottom},
    xshift=-\pgfkeysvalueof{/tikz/afit/left}+\pgfkeysvalueof{/tikz/afit/right},
    yshift=-\pgfkeysvalueof{/tikz/afit/bottom}+\pgfkeysvalueof{/tikz/afit/top}},
    afit/.cd,left/.initial=2pt,right/.initial=2pt,bottom/.initial=2pt,top/.initial=2pt}
    
\tikzstyle{branch}=[{circle,inner sep=0pt,minimum size=0.3em,fill=black}]
\tikzstyle{box} = [draw, dotted, inner xsep=4mm, inner ysep=3mm]
\tikzstyle{every path}=[line width=0.1em]
\tikzstyle{vecArrow} = [thick, decoration={markings,mark=at position
   1 with {\arrow[semithick]{open triangle 60}}},
   double distance=.75pt, shorten >= 5.5pt,
   preaction = {decorate},
   postaction = {draw,line width=.75pt, white,shorten >= 4.5pt}]
\tikzstyle{innerWhite} = [semithick, white,line width=.75pt, shorten >= 4.5pt]
\tikzstyle{normalArrow} = [thick, decoration={markings,mark=at position
   1 with {\arrow{Triangle[length=6pt, width=5.75pt]}}}, preaction = {decorate}, line width=1.25pt, shorten >= 5.5pt]
\tikzstyle{dashedArrow} = [thick, dashed, decoration={markings,mark=at position
   1 with {\arrow{Triangle[length=6pt, width=5.75pt]}}}, preaction = {decorate}, line width=1.25pt, shorten >= 5.5pt]
   
\begin{tikzpicture}[auto, line width=0.1em, font=\Large]
\node[block2, left color=orange!50!white, right color=orange!10!white, at={(0,0)}] (sync) {Synchronization};
\node[block2, left color=lightgreen!50!white, right color=lightgreen!10!white, below=1.5em of sync] (diary) {Spatial \\ Diarization};
\node[block, left color=lightblue!50!white, right color=lightblue!10!white, below=2.75em of diary] (extract) {Guided Source \\Separation};
\node[block, left color=lightblue!50!white, right color=lightblue!10!white, below=1.5em of extract] (mvdr) {Beamforming};
\node[block2, left color=yellow!50!white, right color=yellow!10!white, below=2.75em of mvdr] (asr) {Speech \\ Recognition};
\begin{scope}[on background layer]
 \node[draw, inner sep=1.75em, left color=lightblue!50!white, right color=lightblue!10!white, thick, rectangle, rounded corners=0.1cm, fit={(extract) (mvdr)}] (overall_extract){};
\end{scope}
\node[rotate=90, at={(-6em, -13.em)}] {Source Extraction};

\draw[normalArrow] ($(sync.north) + (0, 1.5em) $)  -- (sync.north);
\draw[normalArrow] (sync) -- (diary);
\draw[normalArrow] (diary) -- (extract);
\draw[normalArrow] (extract) -- (mvdr);
\draw[normalArrow] (mvdr) -- (asr);



\draw[normalArrow] (asr.south) -- ($(asr.south) - (0, 1.5em) $);

\node[block1, at={(20em, -1.5em)}] (gcc) {GCC-PhaT};
\node[block1, below=1.5em of gcc] (peak) {Peak \\ Detection};
\node[block1, below=4em of peak] (combination) {TDOA \\ Combination};
\node[block1, below=1.5em of combination] (count) {Speaker \\ Counting};
\node[block1, below=1.5em of count] (cluster) {TDOA \\ Clustering};
\node at ($(gcc.north) + (0, 3.5em) $)[circle,fill,inner sep=1.5pt](dot){};

\begin{scope}[on background layer]
\node[draw, left color=lightgreen!50!white, right color=lightgreen!10!white, fit margins={left=1em,right=1.7em,bottom=.5em,top=.5em},  thick, rectangle, rounded corners=0.1cm, fit={(dot) (cluster)}](overall) {};
\node[draw, inner sep=1em, left color=lightgreen!45!white, right color=lightgreen!15!white, thick, rectangle, rounded corners=0.1cm, fit={(gcc) (peak)}, xshift=.5cm, yshift=-.5cm](overall_diary5) {};
\draw[normalArrow]($(gcc.north) + (0, 3.5em) $)  -- ($(gcc.north) + (0.5cm, 3.0em) $)-| (overall_diary5.center);
\draw[normalArrow] (overall_diary5.center) -- ($(combination.north) + (.5cm, 0) $);

\node[draw, inner sep=1em, left color=lightgreen!45!white, right color=lightgreen!15!white, thick, rectangle, rounded corners=0.1cm, fit={(gcc) (peak)}, xshift=0.4cm, yshift=-0.4cm](overall_diary4) {};
\draw[normalArrow] (overall_diary4.center) -- ($(combination.north) + (0.4cm, 0) $);
\node[draw, inner sep=1em, left color=lightgreen!45!white, right color=lightgreen!15!white, thick, rectangle, rounded corners=0.1cm, fit={(gcc) (peak)}, xshift=0.3cm, yshift=-0.3cm](overall_diary3) {};
\draw[normalArrow] (overall_diary3.center) -- ($(combination.north) + (0.3cm, 0) $);
\node[draw, inner sep=1em, left color=lightgreen!45!white, right color=lightgreen!15!white, thick, rectangle, rounded corners=0.1cm, fit={(gcc) (peak)}, xshift=0.2cm, yshift=-0.2cm](overall_diary2) {};
\draw[normalArrow] (overall_diary2.center) -- ($(combination.north) + (0.2cm, 0) $);
\node[draw, inner sep=1em, left color=lightgreen!45!white, right color=lightgreen!15!white, thick, rectangle, rounded corners=0.1cm, fit={(gcc) (peak)}, xshift=0.1cm, yshift=-0.1cm](overall_diary1) {};
\draw[normalArrow] (overall_diary1.center) -- ($(combination.north) + (0.1cm, 0) $);

\draw[normalArrow]($(gcc.north) + (0, 3.5em) $)  -- ($(gcc.north) + (0.1cm, 3.4em) $)-| (overall_diary1.center);
\draw[normalArrow]($(gcc.north) + (0, 3.5em) $)  -- ($(gcc.north) + (0.2cm, 3.3em) $)-|(overall_diary2.center);
\draw[normalArrow]($(gcc.north) + (0, 3.5em) $) -- ($(gcc.north) + (0.3cm, 3.2em) $)-| (overall_diary3.center);
\draw[normalArrow]($(gcc.north) + (0, 3.5em) $)  -- ($(gcc.north) + (0.4cm, 3.1em) $)-| (overall_diary4.center);
\node[draw, inner sep=1em, left color=lightgreen!45!white, right color=lightgreen!15!white, thick, rectangle, rounded corners=0.1cm, fit={(gcc) (peak)}](overall_diary) {};
\end{scope}

\draw[normalArrow]($(gcc.north) + (0, 6.em) $) -- (gcc.north);
\draw[normalArrow] (gcc) -- (peak);

\draw[normalArrow] (peak) -- (combination.north);
\draw[normalArrow] (combination) -- (count);
\draw[normalArrow] (count) -- (cluster);
\draw[normalArrow] (cluster.south) -- ($(cluster.south) - (0, 2.5em) $);
\draw[thick, line width=1.25pt, shorten >= 5.5pt] (diary.north east) -- (overall.north west) ;
\draw[thick, line width=1.25pt, shorten >= 5.5pt] (diary.south east) -- (overall.south west) ;

\end{tikzpicture}
    \end{adjustbox}
    \caption{Meeting transcription pipeline}
    \label{fig:overview}
\end{figure}
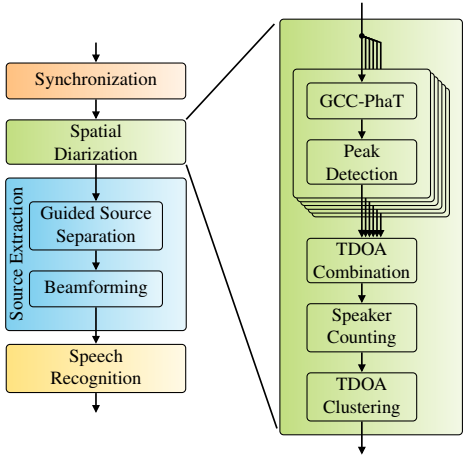
The meeting transcription system, which will be considered in the following, is depicted in \cref{fig:overview}. 
Firstly, the microphone signals are synchronized w.r.t.\ a reference channel. 
To do so, first the \glspl{STO} are compensated for by a correlation-based coarse synchronization~\cite{Araki18,Gburrek22}, which forces the \glspl{TDoA} between the signals to be close to zero at the beginning of the recordings.
Afterwards, the \glspl{SRO} are compensated for via resampling~\cite{Gburrek22}.
The diarization information, which is gathered from \gls{TDoA} information, as well as the estimated \glspl{TDoA} are used to support the extraction of the single speakers' signals from the noisy and reverberant speech mixtures.
Finally, the extracted signals are transcribed.


\section{TDOA-based Diarization}
\label{sec:diarization}
We here propose to cluster frame-wise \gls{TDoA} estimates as representation of the active speakers' positions in order to gather diarization information. 
Therefore, a \gls{TDoA} estimator, that is able to cope with overlapping speech, is introduced.

\subsection{Effect of Asynchronous Recordings}
\label{subsec:tdoa_async}

%
In~\cite{Gburrek22} it was shown that the \gls{TDoA} $\tau_{i,mm'}[\ell]$ between the $m$-th and the $m'$-th channel corresponds to a superposition of the  \gls{TDOF} of the $i$-th speaker's signal between the $m$-th and the $m'$-th channel, a constant offset due to the \gls{STO} and a time-varying \gls{SRO}-induced delay.
Here, $\ell$ denotes the time frame index. 
The  \gls{TDOF} is a characteristic of the $i$-th speaker's position relative to the microphones and, thus, contains spatial information.

The coarse synchronization compensates not only for an \gls{STO} but rather for a combination of \gls{STO}, \gls{SRO}-induced delay and \gls{TDOF}.
Due to this fact the \glspl{TDoA} cannot be mapped to the coordinates of the speakers' positions anymore. 
However, the coarse synchronization affects the \glspl{TDoA} in form of a constant value, which solely depends on the microphone pair. 
Thus, each source position still can be uniquely represented by a vector of all pairwise \glspl{TDoA} $\bm{\tau}_i {=} [\tau_{i,12},\, \tau_{i,13}, \, \dots, \, \tau_{i,M\!-\!1 M}]^T$ after synchronization.

\subsection{Multi-Speaker TDOA Estimation}
\label{subsec:tdoa_est}
As a basis for diarization \gls{TDoA}  vectors are estimated in each time frame (see right half of \cref{fig:overview}).
To this end, the \gls{GCCPhaT}~\cite{Knapp1976} $g_{m m'}(\ell, \lambda)$, with $\lambda$ being the time lag, is firstly estimated for all microphone pairs. 
In order to get more robust \gls{TDoA} estimates, the \gls{GCCPhaT} $g_{m m'}(\ell, \lambda)$ is averaged across $L$ consecutive time frames.
Moreover, the \gls{GCCPhaT} is only calculated on the basis of the frequency range from \SI{125}{Hz} to \SI{3.5}{kHz}, i.e., the frequency range for which speech has significant power.

Since multiple speakers can be active within a time frame, the $C$ time lags $\lambda_c$, belonging to the $C$ highest local maxima of the \gls{GCCPhaT} $g_{m m'}(\ell, \lambda)$, are considered as possible \gls{TDoA} candidates.
Due to the fact that the direct path signal corresponds to a delayed and attenuated version of the source signal, only time lags $\lambda_c$, belonging to positive local maxima, are considered as \gls{TDoA} candidates \cite{scheuing08}.
Furthermore, the local maximum has to be larger than twice the standard deviation of the \gls{GCCPhaT}, which is calculated w.r.t.\ the time lag $\lambda$ for the $\ell$-th time frame. 

Afterwards, the pairwise \gls{TDoA} candidates have to be combined to form consistent \gls{TDoA} vectors.
All elements of a consistent \gls{TDoA} vector have to fulfill the cyclic consistency condition, i.e., in case of three microphones $m$, $n$ and $o$
\begin{align}
    \tau_{mn} - \tau_{mo} +  \tau_{on} \leq \tau_\text{th}, 
    \label{eq:cyclic}
\end{align}
has to be fulfilled, where $\tau_\text{th}$ is a small value of a few samples.
Since we do not check for exact equality to zero in~\eqref{eq:cyclic}, additional valid \gls{TDoA} vectors, e.g., stemming from multi-speaker ambiguities or echos, are possible.
Here, we tackle this issue by utilizing the fact that speaker positions of equal \gls{TDoA} lie on a hyperboloid and the speakers' positions are associated with the point of intersection of the hyperboloids belonging to the different microphone pairs.
Moving along the hyperboloid  of equal \gls{TDoA} of one microphone pair, changes the points of intersection so that the \glspl{TDoA} of all other microphone pairs have to change.
Hence, at maximum  one element is allowed to be equal for two \gls{TDoA} vectors.
In case of multiple \gls{TDoA} vectors having more than one common element, only the \gls{TDoA} vector with the largest \gls{SRPPhaT} is kept. 
Thereby, the \gls{SRPPhaT} is efficiently computed from the previously calculated pairwise \glspl{GCCPhaT} $g_{m m'}(\ell, \lambda)$. 

Finally, the number of  speakers being active within a time frame is determined.
To decide whether there is speech, an energy-based \gls{VAD} is utilized.
In case of speech activity the \gls{TDoA} vector with the largest \gls{SRPPhaT} is considered to belong to an active speaker. 
In addition to that, the \gls{SRPPhaT} is used to decide whether multiple speakers are active.
Additional \gls{TDoA} vectors and, thus, additional speakers for a time frame are considered if the corresponding \gls{SRPPhaT} is larger than the mean of the largest \gls{SRPPhaT} value per frame minus twice their standard deviation.  

\subsection{TDOA Clustering}
\label{subsec:tdoa_cluster}
Diarization information  is gathered by clustering the estimated frame-wise \gls{TDoA} vectors.
First, temporally local clusters, corresponding to speaker activity information approximately at utterance-level, are formed. 
These temporally local clusters are determined via a leader-follower clustering~\cite{vijaya04}.
Thereby, the \gls{TDoA} vector of the most recent frame within a cluster becomes its new leader. 
The temporal locality of the clusters is forced by considering only \gls{TDoA} vectors which do not lie more than \SI{1}{s} in the past as possible leaders.
We use the maximum of the element-wise absolute difference between two \gls{TDoA} vectors as clustering metric.

Subsequently, a single-linkage clustering~\cite{gower69} is employed to obtain the global diarization information from the temporally local clusters.
To this end, the temporally local clusters are represented by the element-wise median of the \gls{TDoA} vectors of their cluster members and the \gls{MSD} between the \gls{TDoA} vectors is used as clustering metric.
The clustering is aborted when the \gls{MSD} is larger than a certain threshold to address outlier \gls{TDoA} vectors.

The final clustering result often contains more clusters than there are speakers.
These clusters mostly belong to \gls{TDoA} vectors which correspond to a combination of direct path  \glspl{TDoA} and \glspl{TDoA} of early reflections or a combination of direct path  \glspl{TDoA} of multiple speakers.
To mitigate these influences, we first sort the estimated speakers' activities by the amount of frames with activity.
If a cluster with a smaller amount of activity intersects more than \SI{50}{\%} with a cluster with a larger amount of activity and more than one element of the \gls{TDoA} vectors of both clusters match each other (see hyperboloild property of \gls{TDoA} vectors described above), the cluster with the smaller amount of activity is discarded.
After all, a dilation and an erosion filter are applied to the estimated activities to smooth the activity estimates~\cite{Boeddeker22}.

\section{Source Extraction}
\label{sec:extraction}
As shown in \cref{fig:overview} mask-based beamforming is utilized to extract the single speakers' signals.
The masks, which are used to calculate the beamformer coefficients, are estimated via a spatial mixture model using the \gls{TDoA}-based diarization information as guide.

\subsection{Guided Source Separation}
\label{subsec:gss}
A time-frequency mask for each speaker and an additional mask for noise are estimated using \gls{GSS}.
In the \gls{GSS} framework, the \gls{TDoA}-based diarization is employed to force the class posterior probability of a spatial mixture model, i.e., the time-frequency masks, to be zero when the corresponding speaker is not active.
In contrast to the original \gls{GSS} method from~\cite{PB2018CHiME5} we here use a \gls{cACGMM}~\cite{Ito2016cACGMM} with time-dependent instead of frequency-dependent mixture weights~\cite{Ito2013permutation} as spatial mixture model.
Since the segmentation needed for \gls{GSS}, which is given by the \gls{TDoA}-based diarization, may also contain segments whose length is underestimated, a context of $\pm \SI{5}{s}$ and additional non-guided \gls{EM} iterations, that follow the guided \gls{EM} iterations, are utilized. 

One way to employ the \gls{TDoA}-based diarization information for initialization of the spatial mixture model is to broadcast the speakers' activities over all frequencies as in the original implementation of \gls{GSS}.
We here propose to utilize the estimated \gls{TDoA} vectors to derive an initial time-frequency mask for each source.
Therefore, a  steering vector based \gls{MVDR} beamformer\cite{haykin02} per speaker is derived from the \gls{TDoA} vectors, assuming anechoic signal propagation. 
The \glspl{SCM} of the interference are calculated as sum of the outer products of the steering vectors of all possibly interfering speakers.
Afterwards the \gls{MVDR} beamformers are applied in the \gls{STFT} domain.
Assuming W-disjoint orthogonality of speech~\cite{yilmaz04} each time-frequency bin is assigned to the mask of the active speaker whose beamformer has the largest output power.

Finally, the method from~\cite{yang19} is used to identify the time-frequency bins which are dominated by a single speaker: 
The \gls{SCM} of the microphone signals is estimated for each time-frequency bin based on a short temporal and frequency context.
Afterwards, the ratio of the largest and the second-largest eigenvalue of the \glspl{SCM} is compared to a certain threshold.
If the  largest eigenvalue is significantly larger than  the second-largest eigenvalue, the time-frequency bin is assumed to be dominated by a single speaker.
All time-frequency bins which are not dominated by a single speaker are assigned to the initial noise mask.



\subsection{Beamforming}
\label{subsec:mvdr}
We utilize an \gls{MVDR} beamformer in the formulation of~\cite{Souden2010MVDR, erdogan16} to extract the signals of the single speakers.
Therefore, we first re-segment  the segments used for \gls{GSS} based on the target speakers' activities, which are calculated from the estimated prior probabilities of the spatial mixture model as described in \cite{Boeddeker22}.
The beamforming coefficients are calculated for each resulting segment, defined by continuous activity of the target speaker, whose signal should be extracted.
The \gls{SCM} of the target speaker is calculated via
\begin{align}
    \bm{\Phi}_i(k) = \frac{1}{|\mathcal{T}_i|} \sum_{\ell \in \mathcal{T}_i} \gamma_i^2(\ell,k){\cdot} \bm{Y}(\ell, k){\cdot} \bm{Y}^H(\ell, k),
\end{align}
with $\mathcal{T}_i$ corresponding to the set of time frames, which belong to the segment, $\gamma_i(\ell,k)$ being the time-frequency mask of the target speaker and $\bm{Y}(\ell, k)$ denoting the vector of stacked \glspl{STFT} of all microphone signals.
The frequency bin index is denoted by $k$.

Since the set of active interfering speakers typically varies over time during a segment, we divide the segment into sub-segments, whose boundaries are given by the change points of the interfering speakers' activities.
For each subsegment new beamformer coefficients are calculated based on the interference \gls{SCM} $\bar{\bm{\Phi}}_{i,b}(k)$, which is estimated via
\begin{align}
    \bar{\bm{\Phi}}_{i,b}(k) =  \frac{1}{|\mathcal{T}_{i,b}|} \sum_{\ell \in \mathcal{T}_{i,b}} (1 - \gamma_i(\ell,k))^2{\cdot} \bm{Y}(\ell, k){\cdot} \bm{Y}^H(\ell, k).
\end{align}
Here, $b$ denotes the index of the subsegment and $\mathcal{T}_{i,b}$ the set of time frames, which belong to the $b$-th subsegment.
The reference channel for beamforming is chosen such that the expected \gls{SDR} of the sub-segment, which exhibits the lowest expected \gls{SDR}, is maximized~\cite{erdogan16}.

\section{Experiments}
\label{sec:exp}
\begin{table}[b]
    \setlength{\tabcolsep}{1.3mm}
    \renewcommand{\arraystretch}{1.2}
    \caption{Comparison of the time frame wise initialization by broadcasting the diarization information along all frequencies (T-Init) and the proposed time-frequency bin wise initialization (TF-Init) for different amounts of guided EM iterations followed by one additional non-guided EM iteration. The signals of the smartphones (Pixel6a, Pixel6b, Pixel7, Xiaomi) from the LibriWASN$^{800}$ data set are used.}
    \centering
    \begin{tabular}{ccccccccc} 
    Guided Iter. & Init.  &\multicolumn{7}{c}{ cpWER / \% } \\
    & & 0L & 0S & OV10 & OV20 & OV30 & OV40 & Avg. \\\hline \hline 
    \multirow{2}{*}{1}&T-Init   & 3.33 & 3.30 & 3.58 & 4.00 & 5.15 & 5.03 & 4.17 \\ \cline{2 - 9}
    &TF-Init  & 3.13 & 2.98 & 3.11 & 3.36 & 4.00 & 3.90 & 3.46  \\ \hline  \hline 
    \multirow{2}{*}{2}&T-Init   & 3.36 & 3.10 & 3.37 & 3.56 & 4.41 & 4.28 & 3.74 \\ \cline{2 - 9}
    &TF-Init & 3.11 & 2.93 & 3.18 & 3.41 & 3.88 & 3.76 & 3.42  \\ \hline  \hline 
    \multirow{2}{*}{5}&T-Init  & 3.25 & 2.93 & 3.22 & 3.31 & 3.83 & 3.83 & 3.43 \\ \cline{2 - 9}
    &TF-Init & 3.11 & 2.97 & 3.20 & 3.34 & 3.84 & 3.67 & 3.39  \\ \hline  \hline 
    \multirow{2}{*}{20}&T-Init  & 3.11 & 2.96 & 3.20 & 3.33 & 3.91 & 3.66 & 3.40 \\ \cline{2 - 9}
    &TF-Init & 3.10 & 2.94 & 3.18 & 3.35 & 3.88 & 3.62 & 3.38\\ \hline  \hline 
    \end{tabular}
    \label{tab:init}
\end{table}

For the experiments we utilize the LibriWASN data set.
The LibriWASN data set consists of recordings of replayed meetings with various overlap conditions, including no overlap (0L \& 0S) as well as \SI{10}{\%} (OV10) to  \SI{40}{\%} (OV40) of speech overlap.
Moreover, the data set offers recordings from two different rooms resulting in the subsets LibriWASN$^{200}$ (reverberation time $\text{T}_{60} {\approx} \SI{200}{ms}$) and LibriWASN$^{800}$ ($\text{T}_{60} {\approx} \SI{800}{ms}$ and computer fan noise in background).
The meetings were recorded by an \gls{ASN} consisting of multiple smartphones and  Raspberry Pis, which were equipped with soundcards.

The system proposed in this contribution is completed by the synchronization and \gls{ASR} building blocks from the reference system provided with the LibriWASN data set in~\cite{schmalen23}.
In order to measure the meeting transcription performance, we employ the \gls{cpWER}~\cite{watanabe20b_chime}.

\subsection{Time-Frequency Mask vs.~Time Mask Initialization}
\label{subsec:initialization}

The influence of the different initialization strategies for the spatial mixture model, i.e., time frame wise initialization by broadcasting the diarization information along all frequencies (T-Init) and the proposed time-frequency bin wise initialization (TF-Init), is shown in \cref{tab:init}.
\gls{GSS} with  up to 20   guided \gls{EM} iterations followed by an additional non-guided \gls{EM} iteration is considered.

It can be seen that the proposed time-frequency bin wise initialization is able to outperform the time frame wise initialization.
This especially holds for the subsets with more speech overlap and when only a few \gls{EM} iterations are used.
Moreover, it becomes obvious that the \gls{cpWER} already converges after a few \gls{EM} iterations for the time-frequency bin wise initialization.
Significantly more \gls{EM} iterations are needed for the time frame wise initialization to end up at a similar performance as for the time-frequency bin wise initialization.

\subsection{Guided Source Separation vs.\ Blind Source Separation}
\label{subsec:guided_vs_blind}
\begin{table}[tb]
    \centering
    \caption{Comparison of the blind spatial mixture model from~\cite{schmalen23} and the \gls{TDoA}-based \gls{GSS} system. Clean denotes transcribing the original LibriSpeech utterances, which were replayed to record the LibriWASN data set.}
    \setlength{\tabcolsep}{1.5mm}
    \renewcommand{\arraystretch}{1.5}
    \begin{tabular}{cccccccccc} 
    \multirow{2}{*}{\rotatebox{90}{\footnotesize Data set}}  & \multirow{2}{*}{\rotatebox{90}{\footnotesize Devices}} & System  &\multicolumn{7}{c}{ cpWER / \% } \\
    & & & 0L & 0S & OV10 & OV20 & OV30 & OV40 & Avg. \\\hline\hline
     &  & Clean & 2.92 & 2.61 & 2.60 & 2.48 & 2.61 & 2.43 & 2.59 \\\hline \hline
    \multirow{4}{*}{\centering \rotatebox[origin=c]{90}{\footnotesize LibriWASN$^{200}$}}& \multirow{2}{*}{\rotatebox[origin=c]{90}{\footnotesize \centering \textit{Phones}}} & Blind  & 2.97 & 2.75 & 2.86 & 2.85 & 3.46 & 2.93 & 2.98 \\ \cline{3 - 10}
    &  & Guided & 2.91 & 2.77 & 2.76 & 2.69 & 3.11 & 2.76 & 2.83  \\ \cline{2 - 10} 
    & \multirow{2}{*}{\centering \rotatebox[origin=c]{90}{\footnotesize \textit{All}}} & Blind & 2.86 & 2.74 & 2.84 & 3.73 & 3.23 & 3.07 & 3.10  \\ \cline{3 - 10} 
    &  & Guided& 2.93 & 2.75 & 2.76 & 2.77 & 2.91 & 2.74 & 2.80  \\ \hline \hline
    \multirow{4}{*}{\centering \rotatebox[origin=c]{90}{\footnotesize LibriWASN$^{800}$}} & \multirow{2}{*}{\centering \rotatebox[origin=c]{90}{\footnotesize \textit{Phones}}}& Blind & 3.04 & 3.09 & 3.75 & 4.76 & 7.71 & 6.08 & 4.96\\ \cline{3 - 10} 
    &   & Guided & 3.11 & 2.93 & 3.18 & 3.23 & 3.64 & 3.54 & 3.30 \\ \cline{2 - 10} 
    & \multirow{2}{*}{\centering \rotatebox[origin=c]{90}{\footnotesize \textit{All}}}  & Blind & 3.09 & 2.93 & 3.25 & 4.46 & 3.69 & 3.12 & 3.45 \\ \cline{3 - 10} 
    &  & Guided & 3.00 & 2.88 & 2.96 & 2.82 & 3.08 & 2.85 & 2.93 \\ \hline \hline
    \end{tabular}
    \label{tab:wer}
\end{table}
\cref{tab:wer} compares the meeting transcription performance which can be reached with the proposed  \gls{TDoA}-based \gls{GSS} system to the performance which can be achieved with the blind spatial mixture model of the LibriWASN reference system from~\cite{schmalen23}, which does not utilize external diarization information.
Furthermore, the initialization of the blind spatial mixture model is not able to cope with overlapping speech and the entire meeting is used at once to estimate the parameters of the blind spatial mixture model.
For a fair comparison, the source extraction proposed in this contribution is adopted to the baseline system. 
\gls{GSS} uses five guided \gls{EM} iterations followed by five non-guided \gls{EM} iterations.
In order to investigate the influence of the amount of available channels, we consider the set \textit{Phones} with four channels, stemming from smartphones (Pixel6a, Pixel6b, Pixel7, Xiaomi) and the set \textit{all} with seven channels stemming from the smartphones and three additional Raspberry Pis with soundcards (asnupb2, asnupb4, asnupb7).

In general, the \gls{TDoA}-based \gls{GSS} system  is able to outperform the blind spatial mixture model.
Thereby, the gap in performance is larger under the more challenging conditions of the LibriWASN$^{800}$ data set with more reverberation and noise.
This especially holds for the sub sets with a larger amount of speech overlap. 
In addition to that, it becomes clear that the \gls{TDoA}-based \gls{GSS}-system  profits from more microphones although decent results can already be achieved with the recordings of four smartphones.

\section{Conclusions}
\label{sec:conclusions}
In this contribution we have shown that spatial information in form of \gls{TDoA} information is a powerful source for diarization information when using an ad-hoc \gls{ASN} in a quite static scenario with spatially well separated speakers like a typical meeting.
Thereby, the benefits of the spatial distribution predominate the challenges arising from the ad-hoc nature of the \gls{ASN}, e.g., unknown microphone positions and asynchronous recordings.
For gathering diarization information, we proposed to cluster \gls{TDoA} estimates from a mutli-speaker \gls{TDoA} estimator. 

Experiments on real recordings have shown that source extraction via mask-based beamforming benefits from the derived diarization information and the \gls{TDoA} estimates from which the diarization information is derived. 
On the one hand, a spatial mixture model, which utilizes the \gls{TDoA}-based diarization as guide, outperforms a blind spatial mixture model with state-of-the-art initialization. 
On the other hand, a time-frequency bin wise initialization based on the \gls{TDoA} estimates leads to a faster convergence of the  spatial mixture model compared to a conventional time frame
wise initialization scheme.

\section*{Acknowledgment}
Partially funded by the Deutsche Forschungsgemeinschaft (DFG, German Research Foundation) - Project 282835863.

\bibliographystyle{IEEEtran}
\bibliography{library}
\end{document}